\documentclass[preprintnumbers, floatfix,letterpaper,aps,prd,epsfig,nofootinbib,onecolumn]{revtex4-1}

\usepackage{bm,graphicx,dcolumn,epstopdf,epsf, latexsym, mathbbol, amssymb,amsmath,color,slashed, mathrsfs,mathcomp, simplewick, natbib}
\usepackage[center]{subfigure}
\usepackage{multirow}
\usepackage{makecell}
\usepackage[colorlinks,linkcolor=blue,citecolor=blue,urlcolor=blue]{hyperref}

\begin{document}
	\allowdisplaybreaks
	\newcommand{\bq}{\begin{equation}}
	\newcommand{\eq}{\end{equation}}
	\newcommand{\bqn}{\begin{eqnarray}}
	\newcommand{\eqn}{\end{eqnarray}}
	\newcommand{\nb}{\nonumber}
	\newcommand{\lb}{\label}
	\newcommand{\f}{\frac}
	\newcommand{\p}{\partial}
	\newcommand{\PRL}{Phys. Rev. Lett.}
	\newcommand{\PLB}{Phys. Lett. B}
	\newcommand{\PRD}{Phys. Rev. D}
	\newcommand{\CQG}{Class. Quantum Grav.}
	\newcommand{\JCAP}{J. Cosmol. Astropart. Phys.}
	\newcommand{\JHEP}{J. High. Energy. Phys.}
	\newcommand{\red}{\textcolor{red}}
	\newcommand{\bs}{\boldsymbol}
	
\title{Constraints on the ghost-free parity-violating gravity from Laser-ranged Satellites}
\author{Jin Qiao${}^{a,b,c}$}
\email{qiaojin@pmo.ac.cn}

\author{Zhao Li ${}^{b,c}$}

\author{Ran Ji ${}^{d}$ }

\author{Tao Zhu${}^{e,f}$}
\email{zhut05@zjut.edu.cn}

\author{Guoliang Li${}^{a,b}$}

\author{Wen Zhao${}^{b,c}$}
\email{wzhao7@ustc.edu.cn}
	
\author{Jiming Chen${}^{g}$}

\affiliation
{
${}^{a}$  Purple Mountain Observatory, Chinese Academy of Sciences, Nanjing, 210023, P.R.China\\
${}^{b}$ School of Astronomy and Space Sciences, University of Science and Technology of China, Hefei, 230026, P.R.China\\
${}^{c}$ Department of Astronomy, University of Science and Technology of China, Hefei 230026, P.R.China \\
${}^{d}$ Research School of Physics, Australian National University, Acton, ACT, 2601, Australia \\
${}^{e}$ Institute for Theoretical Physics and Cosmology, Zhejiang University of Technology, Hangzhou, 310032, P.R.China\\
${}^{f}$ United Center for Gravitational Wave Physics (UCGWP), Zhejiang University of Technology, Hangzhou, 310032, P.R.China \\
${}^{g}$ North Information Control Research Academy Group Co. Ltd. China, Nanjing, 211153, P.R.China; \\
}
	
\date{\today}
	
\begin{abstract}

This paper explores the evolutionary behavior of the Earth-satellite binary system within the framework of the ghost-free parity-violating gravity and the corresponding discussion on the parity-violating effect from the laser-ranged satellites. For this purpose, we start our study with the Parameterized Post-Newtonian (PPN) metric of this gravity theory to study the orbital evolution of the satellites in which the spatial-time sector of the spacetime is modified due to the parity violation. With this modified PPN metric, we calculate the effects of the parity-violating sector of metrics on the time evolution of the orbital elements for an Earth-satellite binary system. We find that among the five orbital elements, the parity violation has no effect on the semi-latus rectum, while the eccentricity and ascending node are affected only in a periodic manner. These three orbital elements are the same as the results of general relativity and are also consistent with the observations of the present experiment.
In particular, parity violation produces non-zero corrections to the eccentricity and pericenter, which will accumulate with the evolution of time, indicating that the parity violation of gravity produces observable secular effects. The observational constraint on the parity-violating effect is derived by confronting the theoretical prediction with the observation by the LAGEOS II pericenter advance, giving a constraint on the parity-violating parameter space from the satellite experiments.
 

\end{abstract}

\maketitle
	
\section{Introduction}
\renewcommand{\theequation}{1.\arabic{equation}} \setcounter{equation}{0}

Einstein's theory of general relativity (GR) was proposed over a century ago and has successfully passed a large number of experimental and observational tests, both in weak field limits and strong field regimes. Observations on the laboratory and solar-system scale where gravity is weak and approximately static \cite{Sabulsky2019, hoyle2001, will2014}, or the strong-field, static observations of binary pulsar systems \cite{stairs2003, wex2014}, and even extreme-gravity observations of GWs \cite{LIGOScientific:2020tif} by the LIGO/Virgo Collaboration and the image of black hole captured by the EHT \cite{eht1, eht2}, are all ultimately found to be in agreement remarkably with the predictions of GR. With such impressive experimental and observational support, GR has become a standard formalism for describing the spacetime around gravitational objects. 

On the other hand, symmetry is an essential characteristic of the fundamental theories of modern physics and thus necessary to be tested experimentally and observationally. It is well known early in the 1950s that one of the discrete symmetries, parity, is violated in the weak interaction \cite{lee_yang_1954, wucs1957}. This essentially implies that our nature is parity-violating. One natural question now arises, that is, whether the parity symmetry can be violated in the gravitational sector.  Recently,  there were lots of theoretical models of gravity with parity violation were studied in the literature, including Chern-Simons gravity \cite{cs0, cs1,cs2, cs3, cs4, cs5, Li:2023lqz}, ghost-free scalar-tensor gravity \cite{ghost-free}, the symmetric teleparallel equivalence of GR theory \cite{Conroy}, the parity-violating scalar-nonmetricity \cite{Zhu:2023wci}, Ho\v{r}ava- Lifshitz gravity \cite{horava, horava1}, and the Nieh-Yan modified teleparallel gravity \cite{NY1, NY2}. These models have been proposed mainly to account for the nature of dark energy, dark matter, or quantizing gravity. It is also shown that the parity violation in the gravitational sector is almost inevitable \cite{cs5} in some quantized theories of gravity, such as string theory and loop quantum gravity. These have already stimulated a lot of experimental and observational tests of gravitational parity violation both in the weak field and strong field regimes, for examples, see \cite{Wang:2020cub, Zhao:2019szi, Zhao:2019xmm, Qiao:2019wsh, bhb, double_pulsar, double_pulsar2, seto, cs3, cs4, cs1, smith2008} and references therein. 

When the parity is violated in the gravitational sector, it can lead to a unique correction to the spatial-time components of the spacetime metric. In the weak field limits, this implies that it modifies the gravitomagnetic sector of spacetime in GR. For example, in the (non-)dynamical Chern-Simons (CS) gravity, the effects of the parity violation normally appear in the spatial-time components of a slowly rotating gravitational object\cite{cs_ppn0, cs_ppn1,slow_rotating_cs1, slow_rotating_cs2, cs2, slow_rotating_cs4}. The parity-violating effects, although might be different in different theories, can introduce new contributions to the frame-dragging effects induced by a rotating or a moving object. Therefore, the measurement of the frame-dragging effect in solar system experiments, such as LAGEOS and LARES satellites \cite{Ciufolini:2019ezb} and Gravity Probe B (GPB) missions\cite{Everitt:2011hp}, are expected to constrain the parity-violating effects of specific theories. In the strong field regime, the most direct tests of gravity are the signals of gravitational wave events, especially parity-violating gravity, whose presence affects the propagation and production of gravitational waves. The presence of the parity violation in gravity induces amplitude or/and velocity birefringence in the propagation of GWs \cite{cs3,Li:2021mdp,Nojiri:2019nar,Nojiri:2020pqr,Sulantay:2022sag, Zhu:2023wci,Jenks:2023pmk,Rao:2023doc}. These parity-violating effects have been examined using gravitational wave event data for various precision tests, like searching for birefringence effects or constraints on the parameters of the theory \cite{Wang:2020cub,Wu:2021ndf, Zhu:2023wci, Zhu:2022uoq, Zhu:2022dfq, Niu:2022yhr, Gong:2021jgg, Ng:2023jjt}. In addition, since frame-dragging effects in the periastron precession of the Double Pulsar PSR J0737-3039A/B might be observed in the near future\cite{Kramer:2021jcw}, the current and future observations of the Double Pulsar can also be used to probe the parity violation.

Various parity-violating theories are currently constructed based on different objectives, of which we mainly focus on the study related to the ghost-free parity-violating gravity \cite{ghost-free}. We find that the presence of parity violation in this theory leads to both velocity and amplitude birefringence, where the effect of velocity birefringence is much greater than that of amplitude birefringence for the gravitational wave signal\cite{Qiao:2019wsh,Zhao:2019xmm}. These effects are examined in Ref. \cite{Wang:2020cub} using gravitational wave data and the tightest constraint on the current theoretical parameter is given. By considering the slow motion and weak field approximations of the theory in the PPN framework, we obtain the new curl-type term similar to those in CS theory due to the presence of parity violation \cite{Qiao:2021fwi}. This shows that parity violation does modify the gravitomagnetic sector of spacetime, i.e., it affects the spacetime around moving or rotating objects. Therefore, two excellent solar system experiments, LAGEOS and LARES satellites\cite{Ciufolini:2019ezb} and GPB \cite{Everitt:2011hp}, are available for testing the parity symmetry of gravity. In this paper, we mainly combine the LAGEOS and LARES satellites experiment to explore the parity symmetry of gravity, while the analysis in combination with the GPB experiment has been discussed in Ref.\cite{Qiao:2021fwi}.

This paper is organized as follows. In Sec. II, we briefly introduce the theory of ghost-free parity-violating gravity. In Sec. III, we calculate the orbital elements. In Sec. IV, we constrain the parameter of this theory with the experimental data. The conclusions and discussions are given in Sec. V. Throughout this paper, we have adopted geometric units such that $G=1=c$, and the metric convention is chosen as $(-, +, +, +)$, and Greek indices $(\mu, \nu, \cdots)$ run over 0, 1, 2, 3 and Latin indices $(i, j, k, \cdots)$ run over 1, 2, 3. 

\section{Parity violating sector of metrics}
\renewcommand{\theequation}{2.\arabic{equation}} \setcounter{equation}{0}

The parity violation in gravitational interaction can arise from various beyond-GR theories, for example the ghost-free parity-violating gravity. In this section, we present a brief introduction to the ghost-free parity-violating gravity. The action of the ghost-free parity-violating gravity has the following form 
\bqn\label{action}
\mathcal{S} = \f{1}{16 \pi G} \int d^4 x \sqrt{-g}\left[ R+\mathcal{L}_{\rm PV}+\mathcal{L}_{\vartheta} + \mathcal{L}_{\rm m}\right]\;,
\eqn
where $R$ is the Ricci scalar, $\mathcal{L}_{\rm PV}$ is the Lagrangian which contains parity-violating terms coupled to a scalar field $\vartheta$, $\mathcal{L}_\vartheta$ is the Lagrangian for the scalar field, which is non-minimally coupled to gravity, and $\mathcal{L}_{\rm m}$ denotes the Lagrangian density of the matter field. The parity-violating Lagrangian $L_{\rm PV} $ has different expressions for different theories. The Chern-Simons Lagrangian can be written in the form \cite{cs5}

\bqn
\mathcal{L}_{\rm CS} = \frac{1}{4}\vartheta \;^*R R\;,
\eqn
where
\bqn
\;^*R R\;= \frac{1}{2} \varepsilon^{\mu\nu\rho\sigma} R_{\rho\sigma \alpha\beta} R^{\alpha \beta}_{\;\;\;\; \mu\nu}
\eqn
is the Pontryagin density defined with $\varepsilon_{\rho \sigma \alpha \beta}$ being the Levi-Civit\'{a} tensor defined in terms of the antisymmetric symbol $\epsilon^{\rho \sigma \alpha \beta}$ as $\varepsilon^{\rho \sigma \alpha \beta}=\epsilon^{\rho \sigma \alpha \beta}/\sqrt{-g}$ and $\vartheta$ is a function of the scalar field $\phi$. 
However, this theory has a higher-derivative field equation, which induces the dangerous Ostrogradsky ghosts. For this reason, CS modified gravity can only be treated as a low-energy truncation of a fundamental theory. To cure this problem, the extension of CS gravity by considering the terms that involve the derivatives of a scalar field is recently proposed in \cite{ghost-free}. The action is generalized in this paper by including the first and second derivatives of the scalar field: $\phi_{\mu} = \nabla_{\mu}\phi$ and $\phi_{\mu\nu} = \nabla_{\mu}\phi_{\nu}$.

$\mathcal{L}_{\rm PV1}$ is the Lagrangian containing the first derivative of the scalar field, which is given by
\bqn\lb{LMPV1}
\mathcal{L}_{\rm PV1} &=& \sum_{\rm A=1}^4  a_{\rm A}(\phi, \phi^\mu \phi_\mu) L_{\rm A},\label{lv1}\\
L_1 &=& \varepsilon^{\mu\nu\alpha \beta} R_{\alpha \beta \rho \sigma} R_{\mu \nu}{}^{\rho}{}_{\lambda} \phi^\sigma \phi^\lambda,\nonumber\\
L_2 &=&  \varepsilon^{\mu\nu\alpha \beta} R_{\alpha \beta \rho \sigma} R_{\mu \lambda }^{\; \; \;\rho \sigma} \phi_\nu \phi^\lambda,\nonumber\\
L_3 &=& \varepsilon^{\mu\nu\alpha \beta} R_{\alpha \beta \rho \sigma} R^{\sigma}_{\;\; \nu} \phi^\rho \phi_\mu,\nonumber\\
L_4 &=&  \varepsilon^{\mu\nu\rho\sigma} R_{\rho\sigma \alpha\beta} R^{\alpha \beta}_{\;\;\;\; \mu\nu} \phi^\lambda \phi_\lambda,\nonumber
\eqn
where $a_{\rm A}$ are a priori arbitrary functions of $\phi$ and $\phi^\mu \phi_\mu$. In order to avoid the Ostrogradsky modes in the unitary gauge (where the scalar field depends on time only), it is required that $4a_1+2 a_2+a_3 +8 a_4 =0$. With this condition, the Lagrangian in Eq.(\ref{lv1}) does not have any higher-order time derivative of the metric, but only higher-order space derivatives.

One can also consider the terms which contain second derivatives of the scalar field. Focusing on only these that are linear in Riemann tensor and linear/quadratically in the second derivative of $\phi$, the most general Lagrangian $\mathcal{L}_{\rm PV2}$ is given by 
\bqn\lb{LMPV2}
\mathcal{L}_{\rm PV2} &=& \sum_{\rm A=1}^7 b_{\rm A} (\phi,\phi^\lambda \phi_\lambda) M_{\rm A},\\
M_1 &=& \varepsilon^{\mu\nu \alpha \beta} R_{\alpha \beta \rho\sigma} \phi^\rho \phi_\mu \phi^\sigma_\nu,\nonumber\\
M_2 &=& \varepsilon^{\mu\nu \alpha \beta} R_{\alpha \beta \rho\sigma} \phi^\rho_\mu \phi^\sigma_\nu, \nonumber\\
M_3 &=& \varepsilon^{\mu\nu \alpha \beta} R_{\alpha \beta \rho\sigma} \phi^\sigma \phi^\rho_\mu \phi^\lambda_\nu \phi_\lambda, \nonumber\\
M_4 &=& \varepsilon^{\mu\nu \alpha \beta} R_{\alpha \beta \rho\sigma} \phi_\nu \phi_\mu^\rho \phi^\sigma_\lambda \phi^\lambda, \nonumber\\
M_5 &=& \varepsilon^{\mu\nu \alpha \beta} R_{\alpha \rho\sigma \lambda } \phi^\rho \phi_\beta \phi^\sigma_\mu \phi^\lambda_\nu, \nonumber\\
M_6 &=& \varepsilon^{\mu\nu \alpha \beta} R_{\beta \gamma} \phi_\alpha \phi^\gamma_\mu \phi^\lambda_\nu \phi_\lambda, \nonumber\\
M_7 &=& (\nabla^2 \phi) M_1,\nonumber
\eqn
with $\phi^{\sigma}_\nu \equiv \nabla^\sigma \nabla_\nu \phi$. Similarly, in order to avoid the Ostrogradsky modes in the unitary gauge, the following conditions should be imposed: $b_7=0$, $b_6=2(b_4+b_5)$ and $b_2=-A_*^2(b_3-b_4)/2$, where $A_*\equiv \dot{\phi}(t)/N$ and $N$ is the lapse function. In this paper, we consider a general scalar-tensor theory with parity violation, which contains all the terms mentioned above. So, the parity-violating term in Eq.(\ref{action}) is given by
\bqn
\mathcal{L}_{\rm PV} = \mathcal{L}_{\rm CS} + \mathcal{L}_{\rm PV1} + \mathcal{L}_{\rm PV2}.
\eqn
Therefore, the CS modified gravity in \cite{cs5}, and the ghost-free parity-violating gravities discussed in  \cite{ghost-free} are all the specific cases of this Lagrangian. The coefficients $\vartheta$, $a_{\rm A}$ and $b_{\rm A}$ depend on the scalar field $\phi$ and its evolution.

The modified field equation can be obtained by variation of the action with respect to the metric $g_{ab}$, which yield
\bqn
R_{ab} - \frac{1}{2}g_{ab} R+  \frac{1}{\sqrt{-g}} \frac{\delta (\sqrt{-g}   \mathcal{L}_{\rm PV} )}{\delta g^{\mu\nu}} 
= - \frac{2}{\sqrt{-g}} \frac{\delta (\sqrt{-g} \mathcal{L}_{\rm m} + \sqrt{-g} \mathcal{L}_{\phi } )}{\delta g^{\mu\nu}}. \lb{GR}
\eqn
In Ref.\cite{Qiao:2021fwi}, we have obtained the specific form of this modified field equation and derived the PPN metrics of the perturbative field equation in the PPN framework.
The full PPN metrics of this theory are given by
\bqn\lb{ppe1}
g_{00} &=& -1+ 2U -2U^2 + \Psi + \mathcal{O}(6), \nb\\
g_{0i} &=& -\frac{7}{2} V_i - \frac{1}{2}W_i + 2f'(\nabla \times V)_i +\mathcal{O}(5), \nb\\
g_{ij} &=& (1+2U) \delta_{ij} + \mathcal{O}(4),
\eqn
where $f  = \vartheta + (- 2a_2 + a_3 -8 a_4)\phi'^2$, $\Psi = 4\Phi_1 + 4\Phi_2 + 2\Phi_3 + 6\Phi_4$, the PPN potentials $U,~\Phi_1,~\Phi_2,~\Phi_3,~\Phi_4,~V_i,~W_i$ are given by Eqs. (\ref{ppe2}-\ref{SPchi}) in Appendix A. Note that the parity-violating terms in the gravity only contribute to the $g_{0i}$ sector of the metric. 
Therefore, for this metric with parity-violating corrections, it can be tested in conjunction with solar system experiments, such as GPB, LAGEOS, LAGEOS II, and LARES satellites.
In this paper, we will consider the effect of parity violation on the evolution of particle orbits in such a binary system of the Earth and its satellite with a revolution velocity around the Sun.

\section{Procession rate of the orbital elements}
\renewcommand{\theequation}{3.\arabic{equation}} \setcounter{equation}{0}

Now we consider a particle moving in the above metric (\ref{ppe1}). The trajectory of the particle follows time-like geodesic, which is governed by
\bqn
\frac{d^2x^{\mu}}{d \tau^2} = - \Gamma^{\mu}_{\nu \lambda} \frac{dx^\nu}{d\tau} \frac{d x^\lambda}{d\tau}, 
\eqn
where $x^\mu=( t, \; x, \; y, \; z)$ is the coordinates of the test particle, $\tau$ is the proper time, and $\Gamma^{\mu}_{\nu \lambda} $ is the Christoffel connection of the spacetime. We are interested in the acceleration of the particle, and we find
\bqn
\frac{d^2x^i}{dt^2} &=&   - \Gamma^i_{00} - 2 \Gamma^{i}_{0j}v^j - \Gamma^i_{jk}v^j v^k + \Gamma^0_{00}v^i  + 2\Gamma^0_{0j}v^i v^j + \Gamma^0_{jk}v^i v^j v^k, 
\eqn 
where  $v^ i\equiv dx^i/dt \ll1$. Eq.(\ref{ppe1}) has given the metric form, in order to simplify the later calculation we define a new vector $g_{0i}=\zeta_i$.
In the weak field limit, it is a good approximation to consider $v^i \ll1 $ and we can retain the leading order and ignore the higher-order terms.
Then the acceleration of the particle takes the form as
\bqn
\frac{d^2 {\bs r}}{dt^2} &=& \nabla U   - 2U \nabla U + {\bs v} \times \left( \nabla\times {\bs \zeta} \right) 
 -3 (\nabla U\cdot {\bs v}) {\bs v} + \nabla U  {\bs v}^2 + \mathcal{O}(5).
\eqn
We also neglected the contribution of $\Psi$ in the above equation, which is not corrected in this theory.
Note that the velocity of the satellite in the Earth-satellite binary system contains two contributions: ${\bs v} = {\bs v}_s + {\bs v}_{\oplus}$, where ${\bs v}_s$ indicates the orbital velocity of the satellite around the Earth and ${\bs v}_{\oplus}$ indicates the velocity of the motion of the binary system around the Sun.

For an orbiting, rotating, almost spherical gravitational object, such as the Earth, the PPN potential can be written as \cite{will2018th}
\bqn
U &=& \frac{M_{\oplus}}{r}, \nb\\
{\bs V} &=& \frac{M_{\oplus}}{r} {\bs v}_{\oplus} + \frac{1}{2} \left( \frac{{\bs J}_{\oplus}}{r^3} \times {\bs x} \right),\nb\\
{\bs W} &=& \frac{M_{\oplus}}{r} ({\bs n} \cdot {\bs v}_{\oplus} ){\bs n}+ \frac{1}{2} \left( \frac{{\bs J}_{\oplus}}{r^3} \times {\bs x} \right),
\eqn
where ${\bs n} \equiv {\bs r}/r$, $M_{\oplus}$ and ${\bs J}_{\oplus}$ denote the mass and angular momentum of the Earth, respectively.
We now treat all the contributions to the acceleration except the Newtonian two-body acceleration as perturbations, i.e.,
\bqn
\frac{d^2 {\bs r}}{dt^2} = - \frac{M_{\oplus}}{r^2} \frac{\bs r}{r}  + {\bs F_1} + {\bs F_2} + {\bs F_3}, 
\eqn
where ${\bs F_1},~{\bs F_2}, ~{\bs F_3}$ denote the perturbed force to the binary system which are given by
\bqn\lb{f123}
{\bs F_1} &=& 4\f{M^2_{\oplus}}{r^3} {\bs n} +3\f{M_{\oplus}}{r^2}  ({\bs n}\cdot {\bs v}_{s}) {\bs v}_{s} - M_{\oplus}\f{{\bs n}}{r^2}   {\bs v}_{s}^2 
                          - \f{6  }{r^3}({\bs v}_{s} \times {\bs n}   ) ({\bs n} \cdot{\bs J_{\oplus}}) + \f{2 }{r^3} {\bs v}_{s}\times {\bs J_{\oplus}}   , \nb\\ 
{\bs F_2} &=&    { - \f{M_{\oplus}}{r^2} [ ({\bs n}\cdot  {\bs v}_{s}) {\bs v}_{\oplus} -3 ({\bs n}\cdot {\bs v}_{\oplus} ) {\bs v}_{s}+  ({\bs n}\cdot  {\bs v}_{\oplus}) {\bs v}_{\oplus} ]} \nb\\
                          &&{+ M_{\oplus}\f{{\bs n}}{r^2}  (   2 {\bs v}_{\oplus} \cdot  {\bs v}_{s} -3 {\bs v}^2_{\oplus} )}
                            {- \f{6  }{r^3}( {\bs v}_{\oplus} \times {\bs n} ) ({\bs n} \cdot{\bs J_{\oplus}}) +  \f{2 }{r^3}( {\bs v}_{\oplus} \times{\bs J_{\oplus}} )}, \nb\\ 
{\bs F_3} &=&    -2 f'  \f{M_{\oplus}}{r^3}  {\bs v}_{s} \times {\bs v}_{\oplus} +6 f' \f{M_{\oplus}}{r^3} ( {\bs v}_{s} \times {\bs n}) ({\bs n}\cdot {\bs v}_{\oplus} ) 
                          +6 f' \f{M_{\oplus}}{r^3} ({\bs v}_{\oplus}  \times {\bs n}) ({\bs n}\cdot {\bs v}_{\oplus} ).
\eqn
Here ${\bs F_1} $ depends on the mass and spin of the Earth, which denotes the perturbative force of the Earth's gravity and spin on the satellite orbit. ${\bs F_2}$  depends not only on the mass and spin of the Earth, but also on the  translational velocity of the Earth, which denotes that the perturbative force contains the contribution of the Sun's gravity. It is important to note that ${\bs F_1}$ and ${\bs F_2} $  are both derived from the orbital force generated in GR theory, only distinguishing the different states of motion of the binary system.
However, ${\bs F_3}$ also depends on an additional parameter $f'$, which indicates the perturbative force generated by the parity-violating terms. We then consider the effects of each of these three perturbative forces on the orbital motion.

In this way, one can define the osculating elliptical orbit of the binary motion as
\bqn
{\bs r} &\equiv& \frac{p}{1+e \cos f} \hat{\bs n},~~~
{\bs v_s} \equiv \dot r \hat{\bs n} + \frac{h}{r} \hat{\bs \lambda}, ~~~
h \equiv  \sqrt{GM_{\oplus}p},
\eqn
where $p$ is the semi-latus rectum of the elliptical motion, $e$ is the orbitial eccentricity, and $f$ is the true anomaly. The unit vectors $\hat{\bs n}$ and $\hat{\bs \lambda}$ denote the radial and transverse directions respectively. With $\hat{\bs n}$ and $\hat{\bs \lambda}$ we can introduce the unit normal vector $\hat {\bs h} = \hat{\bs n} \times \hat{\bs \lambda}$. The unit vectors $\hat{\bs n}$, $\hat{\bs \lambda}$, and $\hat {\bs h}$ can be related to the fixed reference directions ${\bs e_x}$, ${\bs e_y}$, and ${\bs e_z}$ of the coordinate system $(x, y, z)$ via the following transformation, 
\bqn
\hat{\bs n} = \cos f {\bs e_{x}} + \sin f {\bs e_{y}},~~~
\hat{\bs \lambda} =- \sin f {\bs e_{x}} + \cos f {\bs e_{y}},~~~
\hat{\bs h} = {\bs e_{z}},
\eqn
where ($\Omega$, $\iota$, $\omega$) are the longitude of the ascending node, the inclination angle, and the argument of pericentre respectively. 
We need to note that both coordinate frames $({\bs e_x}, {\bs e_y}, {\bs e_z})$ and $(\hat{\bs n}, \hat{\bs \lambda}, \hat {\bs h})$ are established in the satellite orbital plane.
These three quantities are the Euler angles which characterize the rotation between the frame $({\bs e_x}, {\bs e_y}, {\bs e_z})$ and $(\hat{\bs n}, \hat{\bs \lambda}, \hat {\bs h})$. 

In addition, we introduce a fundamental frame with base vectors $({\bs e_X}, {\bs e_Y}, {\bs e_Z})$ and describe the motion of the system in this new frame. We adopt the $Z$-axis as the reference direction parallel to the spin vector ${\bs J_{\oplus}}$ of the Earth and the $X-Y$ plane as the reference plane of the new frame. The fundamental $(X, Y, Z )$ frame can be obtained from the orbital $(x, y, z)$ frame by the following transformation,
\bqn
{\bs e_{X}} &=& [\cos \Omega \cos\omega - \cos \iota \sin\Omega \sin \omega ] {\bs e_{x}}  
 - [\cos \Omega \sin \omega + \cos \iota \sin \Omega \cos \omega ]{\bs e_y}  
  + \sin\iota \sin \Omega {\bs e_z}, \nb\\
{\bs e_{Y}} &=& [\sin \Omega \cos \omega  + \cos\iota \cos \Omega \sin \omega ]{\bs e_x}   
  - [\sin\Omega \sin \omega - \cos\iota \cos\Omega \cos \omega ]{\bs e_y}  
 - \sin\iota \cos \Omega  {\bs e_z}, \nb\\
{\bs e_{Z}} &=& \sin\iota \sin\omega {\bs e_x} + \sin\iota \cos\omega {\bs e_y} + \cos\iota {\bs e_z}.
\eqn
In these two coordinate frames, the vectors ${\bs v_{\oplus}}$ and ${\bs J_{\oplus}}$ can be conveniently given
\bqn
{\bs v_{\oplus}} &=&  v_{\oplus}(\sin \theta \cos \gamma {\bs e}_X +  \sin \theta \sin \gamma {\bs e}_Y +  \cos \theta {\bs e}_Z)\nb\\
                         &=& v_x {\bs e_{x}}+ v_y {\bs e_{y}} + v_z{\bs e_{z}},\nb\\
{\bs J}_{\oplus} &=& J_{\oplus} {\bs e_Z}\nb\\
                         &=& J_x {\bs e_{x}} + J_y{\bs e_{y}} + J_z{\bs e_{z}},
\eqn
where
\bqn\lb{v-xyz}
v_x &=& v_{\oplus} [ \sin\theta\cos\gamma (\cos \Omega \cos \omega - \cos \iota \sin \Omega \sin \omega) 
     + \sin\theta \sin\gamma (\sin \Omega \cos \omega + \cos \iota \cos \Omega \sin \omega)  
     +  \cos \theta (\sin \iota \sin \omega) ],\nb\\
v_y &=& v_{\oplus} [ -\sin\theta\cos\gamma (\cos \Omega \sin \omega + \cos \iota \sin \Omega \cos \omega) 
     - \sin\theta \sin\gamma  (\sin \Omega \sin \omega - \cos \iota \cos \Omega \cos \omega) 
     +   \cos \theta (\sin \iota \cos \omega) ] ,\nb\\
v_z &=&  v_{\oplus}  [ \sin\theta\cos\gamma \sin \iota \sin \Omega -\sin\theta \sin\gamma \sin \iota \cos \Omega +  \cos \theta  \cos \iota ],\nb\\
J_x &=& J_{\oplus} \sin\iota \sin\omega, ~~~ J_y = J_{\oplus} \sin\iota \cos\omega ,~~~ J_z =J_{\oplus} \cos\iota.
\eqn
It should be noted here that two new parameters $\gamma$ and $\theta$ are introduced in the above equation, {where the parameter $\gamma$ denotes the angle between the projection of the velocity $\boldsymbol{v}_{\oplus}$ } in the $X-Y$ plane and the position of the right ascending node and the parameter $\theta$ is between the spin $\boldsymbol{J}_{\oplus}$ and velocity vectors $\boldsymbol{v}_{\oplus}$ of the Earth. The parameters $\gamma$ and $\theta$ are the periodic parameters that vary with time over a period of one year.
With the previous discussion, one can decompose the perturbed force ${\bs F}_i(i=1,2,3)$ along the $\hat{\bs n}$, $\hat{\bs \lambda}$, and $\hat {\bs h}$ as
\bqn
{\bs F}_i = {\cal R}\hat {\bs n} + {\cal S} \hat {\bs \lambda} + {\cal W} \hat {\bs h}.
\eqn

\subsection{ Orbital effects from the perturbation force \texorpdfstring{$\bs{ F}_1$}{} }

In previous calculations, we already know that ${\bs F}_1$ is a perturbation produced by the Earth's gravity and spin on the satellite's orbital motion. We can decompose this perturbation ${\bs F}_1$ into the coordinate system $[\cal R,~S,~W]$, and then bring it into the expression for the orbital elements (\ref{af} - \ref{omegaf}), and finally we can calculate the rate of change for each orbital element. The results of this part of GR have been investigated in detail in many works of literature. Therefore, we do not repeat the calculation, and the final results are given directly below \cite{GNPR}
\bqn\label{GRLT}
\dot \Omega_1 &=& \f{2J_{\oplus}}{a^3(1-e^2)^{3/2}},\nb\\
\dot \omega_1 &=& \f{3M_{\oplus}^{3/2}}{a^{5/2}(1-e^2)} - \f{4J_{\oplus}\cos \iota}{ a^3(1-e^2)^{3/2}}.
\eqn
The results of the theoretical predictions of the general relativity of the secular frame-dragging precession rate $\dot \Omega$ on the right-ascension of the ascending node of the LAGEOS, LAGEOS II, and LARES satellites have been given in  Table. \ref{tableX}.

\subsection{ Orbital effects from the perturbation force \texorpdfstring{$\bs{F}_2$}{} }

Compared with the perturbative force ${\bs F}_1$, ${\bs F}_2$ also depends on the Earth's translational velocity in Eq.(\ref{f123}), that is, the Sun's gravity is involved in the generation of this perturbative force. Decomposing ${\bs F}_2$ into the coordinate system $[\cal R,~S,~W]$ yields
\bqn
{\cal R} &=& \f{M_{\oplus}}{r^2} [ 4 \f{he}{p} \sin f ( v_x\cos f + v_y \sin f )-  ( v_x\cos f + v_y \sin f  )^2 ] \nb\\
                          && + \f{M_{\oplus}}{r^2}  \left[  2 \f{h}{r}(v_y \cos f - v_x \sin f)-3 v^2_{\oplus}  \right] \nb\\
                          &&  + \f{ 2}{r^3} [(v_y J_z -v_zJ_y) \cos f +( v_z J_x -v_xJ_z) \sin f) ], \\
{\cal S} &=&  { - \f{M_{\oplus}}{r^2}  [ \f{he}{p}\sin f (v_y\cos f- v_x\sin f) -3\f{h}{r}(v_x\cos f + v_y\sin f) ]} \nb\\
                          &&  { - \f{M_{\oplus}}{r^2}  [(v_x\cos f+v_y \sin f)(v_y\cos f - v_x\sin f) ]} \nb\\
                          &&  {- \f{6  }{r^3} v_z (J_x\cos f + J_y \sin f)} 
                           {+2  \f{1 }{r^3} [(v_zJ_y - v_y J_z) \sin f +( v_z J_x -v_xJ_z) \cos f) ] } ,  \\
{\cal W} &=& - \f{M_{\oplus}}{r^2} [ \f{he}{p} \sin f v_z  +  (v_x \cos f + v_y \sin f) v_z ] \nb\\
                         &&  {- \f{6  }{r^3} (v_x \sin f - v_y \cos f) (J_x\cos f + J_y \sin f) }
                            {+  \f{2 }{r^3}(v_x J_y - v_y J_x )}  .
\eqn
We insert the above expressions into Eq. (\ref{af} - \ref{omegaf}) and integrate them over a complete orbital period to calculate the secular variation of the orbital elements $p$, $e$, $\iota$, $\omega$, and $\Omega$. Through tedious calculations, we obtain
 \bqn\label{mqfe}
\f{dp}{df} &=& { -2     \f{e h}{1+e\cos f}\sin f (v_y\cos f- v_x\sin f) + 6 h (v_x\cos f + v_y\sin f) } \nb\\
                          &&  { - 2 \f{p}{(1+e\cos f)} (v_x\cos f+v_y \sin f)(v_y\cos f - v_x\sin f) } \nb\\
                          &&  {- \f{8 }{M_{\oplus}} v_z (J_x\cos f + J_y \sin f)} 
                          { - \f{4 }{M_{\oplus}}v_y J_z \sin f - \f{4 }{M_{\oplus}} v_xJ_z \cos f  },   \nb\\
\f{de}{df}  &=& {  4 e\f{h}{p} \sin^2 f ( v_x\cos f + v_y \sin f ) - \sin f ( v_x\cos f + v_y \sin f  )^2  } \nb\\
                          &&{+  2 \f{h}{p} \sin f (1+e\cos f)(v_y \cos f - v_x \sin f) } { -3\sin f  v^2_{\oplus}   }\nb\\
                          && { +2  \f{1+e\cos f }{M_{\oplus}p} \sin f  [(v_y J_z -v_zJ_y) \cos f +( v_z J_x -v_xJ_z) \sin f) ]}   \nb\\
                          &&{ - \f{eh}{p} \f{2\cos f + e(1+\cos^2 f)}{(1+e\cos f)}  \sin f (v_y\cos f- v_x\sin f) } \nb\\
                          &&{ + 3\f{h}{p}[2\cos f + e(1+\cos^2 f)](v_x\cos f + v_y\sin f) } \nb\\
                          &&  { - \f{2\cos f + e(1+\cos^2 f)}{(1+e\cos f)} (v_x\cos f+v_y \sin f)(v_y\cos f - v_x\sin f) } \nb\\
                          &&  {-2 \f{2\cos f + e(1+\cos^2 f) }{pM_{\oplus}}  (2v_z J_x\cos f + 2v_z J_y \sin f + v_y J_z \sin f  + v_xJ_z \cos f)}, \nb\\
\f{d\iota}{df} &=& { -   \f{eh}{p} \f{\sin  (\omega+ f)}{(1+e\cos f)} \sin f v_z  - \f{\sin  (\omega+ f)}{(1+e\cos f)} (v_x \cos f + v_y \sin f) v_z } \nb\\
                          &&  {- \f{6 \sin (\omega+ f) }{pM_{\oplus}} (v_x \sin f - v_y \cos f) (J_x\cos f + J_y \sin f) } {+2  \f{\sin (\omega+ f) }{pM_{\oplus}}(v_x J_y - v_y J_x )},  \nb\\
\sin \iota  \f{d \Omega}{df} &=&  \f{he}{p} \f{\sin  (\omega+ f)}{(1+e\cos f)}\sin f v_z  - \f{\sin  (\omega+ f)}{(1+e\cos f)} (v_x \cos f + v_y \sin f) v_z \nb\\
&&- \f{6 \sin (\omega+ f)}{pM} (v_x \sin f - v_y \cos f) (J_1\cos f + J_2 \sin f)   +2  \f{\sin (\omega+ f)}{pM}(v_x J_2 - v_y J_1 ),  \nb\\
 \f{d \omega}{df} &=& {    -2 \f{h}{p} \f{\cos f}{(1+e\cos f) }  \sin f ( v_x\cos f + v_y \sin f )+  \f{1}{e}  \f{\cos f}{(1+e\cos f) } ( v_x\cos f + v_y \sin f  )^2 } \nb\\
                          &&{-  2\f{h}{p}  \f{\cos f \sin f }{(1+e\cos f)}  (v_x\cos f+ v_y \sin f) -   \f{2h}{ep} \cos f(v_y \cos f - v_x \sin f)}\nb\\
                          &&{ + 3  \f{1}{e}  \f{\cos f}{(1+e\cos f)}v^2_{\oplus}   }
                           { - 2 \f{1}{epM_{\oplus}}  \cos f[(v_y J_z -v_zJ_y) \cos f +( v_z J_x -v_xJ_z) \sin f) ]}   \nb\\
         &&  { -  \f{h}{p} \f{2+e\cos f}{(1+e\cos f)} \sin^2 f (v_y\cos f- v_x\sin f)   } 
                          { + 3 \f{eh}{p}(2+e\cos f) \sin f (v_x\cos f + v_y\sin f) ]} \nb\\
                          &&  { -  \f{1}{e}  \f{2+e\cos f}{(1+e\cos f) }\sin f  [(v_x\cos f+v_y \sin f)(v_y\cos f - v_x\sin f) ]} \nb\\
                          &&  {-  \f{6}{epM_{\oplus}} (2+e\cos f)\sin f  v_z (J_x\cos f + J_y \sin f)} \nb\\
                          && {+2  \f{1}{epM_{\oplus}}  (2+e\cos f)\sin f [- (v_y J_z -v_zJ_y) \sin f +( v_z J_x -v_xJ_z) \cos f) ] } \nb\\
         && { +     \f{eh}{p} \cot\iota \f{\sin(\omega+f)}{(1+e\cos f)} \sin f v_z  +  \cot\iota \f{\sin(\omega+f)}{(1+e\cos f)} (v_x \cos f + v_y \sin f) v_z } \nb\\
                          &&  {+  \cot\iota \f{6 \sin(\omega+f) }{pM_{\oplus}} (v_x \sin f - v_y \cos f) (J_x\cos f + J_y \sin f) } {- 2 \cot\iota \f{\sin(\omega+f) }{pM_{\oplus}}(v_x J_y - v_y J_x )}.    \nb\\
\eqn
The above calculations show that the contribution of the perturbation force ${\bs F}_2$ to the orbital elements is not zero. This part of the equation for the orbital elements has some expressions that depend on the combination of eccentricity $e$ and trigonometric functions, making it difficult to give a concise form of the integral.
To compare with the effects produced by the perturbative force ${\bs F}_1$, we choose only the orbital element $\dot \Omega$ for the magnitude comparison. The relevant parameters have been given in Table. \ref{table} and substituted into the orbital element $\dot \Omega$  to obtain
\bqn
\dot \Omega_{2} \sim 10^{-4} {\rm mas/yr} \ll \dot \Omega_{1} \sim 10^{2} {\rm mas/yr}.
\eqn
This result indicates that the contribution of the perturbation force ${\bs F}_2$ to the orbital elements is negligible compared to the perturbation force ${\bs F}_1$.

\subsection{ Orbital effects from the perturbation force \texorpdfstring{$\bs{ F}_3$}{} }

Compared to the perturbative forces ${\bs F}_1$ and ${\bs F}_2$, each of ${\bs F}_3$ depends on the parameter $f'$. Thus ${\bs F}_3$ is the perturbative force of which we are concerned for the orbital motion produced by the parity-violating correction. Similarly, we decompose ${\bs F}_3$ into the coordinate system $[\cal R,~S,~W]$ to obtain
\bqn\label{f3rsw}
{\cal R} &=& -2 f'  \f{M_{\oplus}h}{r^4}   v_z, \nb\\
{\cal S} &=& 2 f'  \f{M_{\oplus}}{r^3} \f{he}{p}  v_z \sin f +6 f' \f{M}{r^3} v_z (v_x \cos f + v_y \sin f) , \nb\\
{\cal W} &=& -2 f'  \f{M_{\oplus}}{r^3} \f{he}{p}  (v_y \cos f - v_x \sin f) \sin f 
     +8 f' \f{M_{\oplus}h}{r^4} (v_x \cos f + v_y \sin f ) \nb\\
    && +6 f' \f{M_{\oplus}}{r^3}  ( v_x \sin f - v_y \cos f) (v_x \cos f + v_y \sin f),
\eqn
where the forms for $v_x$, $v_y$ and $v_z$ have been given in Eq.(\ref{v-xyz}).
We are then at a position to calculate the secular variation of the orbital elements $p$, $e$, $\iota$, $\omega$, and $\Omega$ by inserting the above expressions into (\ref{af} - \ref{omegaf}), and integrating over a complete satellite orbital cycle. Note that the integration process here is chosen over a satellite orbital period, during which the variation of the parameters $\gamma$ and $\theta$ of the velocity component $v_{x,y,z}$ is very small compared to their period of one year. Therefore, in this process we can treat these two parameters as constants. It is straightforward to obtain,
\bqn\lb{eesc}
\dot p_3 &=& 0, \nb\\
\dot e_3 &=& 6 f' \sqrt{\f{GM_{\oplus}}{a^3}} \f{1}{p} v_x v_z , \nb\\
\dot \iota_3 &=&  4 f'  \sqrt{\f{GM_{\oplus}}{a^3}} \f{h}{p^2}   (v_x \cos \omega - v_y \sin \omega )   ,  \nb\\
\dot \Omega_3 &=&  4 f' \sqrt{\f{GM_{\oplus}}{a^3}}  \f{h}{p^2}   (v_x  \sin \omega + v_y \cos \omega )    ,  \nb\\
\dot \omega_3 &=&  3 f' \sqrt{\f{GM_{\oplus}}{a^3}}  \f{ h}{p^2 }   v_z - 4 f' \sqrt{\f{GM_{\oplus}}{a^3}}  \f{h }{ p^2 } \cot\iota  (v_x  \sin \omega  + v_y  \cos \omega) { + 6 f'\sqrt{\f{GM_{\oplus}}{a^3}} \f{1}{ep} v_y v_z } .
\eqn
{From the expression above we can see that the last four of the orbital elements are non-zero, which indicates that the parity-violating terms all produce corrections to these orbital elements. These results are very similar to those in the Lense-Thirring effect, which is due to the rotation of the object and the creation of pseudo-vectors in the perturbative force  ${\bs F}_1$, that is, the last two terms in expression (\ref{f123}). These terms also lead to corrections in the orbital elements for eccentricity, inclination, ascending node and pericenter, of which only the ascending node and pericenter are true secular effects\cite{Mashhoon:1984fj}, which are the terms associated with the spin $J_{\oplus}$ in expression (\ref{GRLT}). We can see from expression (\ref{f123}) that the perturbative force  ${\bs F}_3$ is a pseudo-vector, and all three of its components $[R, S, W]$ (\ref{f3rsw}) in the coordinate system are pseudo-scalars. Ultimately the corrections to the orbital elements by this perturbation force ${\bs F}_3$ also produce non-zero inclination, eccentricity, ascending node, and pericenter, as well as all four orbital elements are pseudoscalars.  However, whether all these corrected orbital elements have real secular effects will be further discussed in the next section.}

In some applications or solar system experiments, it is not easy to separate the variation of $\Omega$ and $\omega$, especially when the inclination angle $\iota$ is very small. In these cases, one in general measures the procession rate of the periastron along the orbit plane, which is typically represented as
\bqn
\dot \varpi &\equiv & \dot \omega_3 + \dot \Omega_3 \cos \iota.
\eqn
For these metrics, it is easy to show that
\bqn
\dot \varpi &=& 3 f' \sqrt{\f{GM_{\oplus}}{a^3}}  \f{ h}{p^2 }   v_z { + 6 f' \sqrt{\f{GM_{\oplus}}{a^3}} \f{1}{ep} v_y v_z }.
\eqn
From the above calculations, we can clearly see that in such a binary system, the parity-violating terms produce non-zero contributions to the satellite's orbital elements. Therefore, these effects can be verified by relevant experiments. In the following, we will discuss these corrections arising under this theory in the context of specific experiments.

\section{Constraints with experimental results from laser-ranged satellites}
\renewcommand{\theequation}{4.\arabic{equation}} \setcounter{equation}{0}

The LAGEOS, LAGEOS II, and LARES satellites are three man-made laser ranged satellites of ASI, the Italian Space Agency. These satellites are designed to measure the frame-dragging effects of the earth's rotation on the orbits of the satellites. Recently, such frame-dragging effects have been measured to about a few present accuracies by using 7 years of the laser ranged data of LARES and 26 years of the laser-ranged data of LAGEOS and LAGEOS II \cite{LARES}. Such a measurement also used the static part and temporal variations of the Earth's gravity field obtained by the space geodesy mission GRACE (NASA) and in particular the static Earth’s gravity field model GGM05S augmented by a model for the 7-day temporal variations of the lowest degree Earth spherical harmonics \cite{LARES}. By introducing a dimensionless coefficient $\mu$ to represent the frame-dragging effect parameter with $\mu=1$ being the value in GR, the authors in \cite{LARES} consider a measurement of the combination 
\bqn\lb{lg123}
&&\mu \dot \Omega_{I} + \mu k_1 \dot \Omega_{II} + \mu k_2 \dot \Omega_{III} \nb\\
&&~~ \simeq \mu (30.68+k_1 31.50 + k_2 118.50) \; {\rm mas/yr} + {\rm errors}, \lb{combination}
\eqn
where $k_1= 0.3448$, $k_2= 0.07291$, and $\dot \Omega_I $, $\dot \Omega_{II} $, and $\dot \Omega_{III}$ which are given in Table. \ref{tableX}, denote the precession rate on the right-ascension of the ascending node of LAGEOS, LAGEOS II, and LARES satellites, respectively. The latest measurement of the above combination leads to \cite{LARES}
\bqn\lb{lgu}
\mu_{\rm meas}-1= (1.5 \pm 7.4)\times 10^{-3} \pm 16\times 10^{-3} \lb{mu}
\eqn
where the first terms within parentheses represent the measurement and its formal error ($0.74\%$, at a $95\%$ confidence level), while the last term represents estimate for the error budget, $1.6\%$. {As pointed out in ref.\cite{LARES}, }this error budget derives from a root-sum-square (RSS) of the main systematic effects that are related to the gravitational and non-gravitational perturbations that act on the orbits of the satellites. It accounts for the errors related to the static field, to ocean tides, to other periodic effects and to the error related to the knowledge of the de Sitter precession. 


Here if we directly adopt the results of Eq.(\ref{eesc}), using the above measurements can formally go to constrain the parity-violationg effect. However, again we need to note that Eq.(\ref{lg123}) and Eq.(\ref{lgu}) are cumulative observations over a long period of time (more than seven years), while Eq.(\ref{eesc}) is the result obtained over one satellite orbital period. If the evolution time we consider is a long-term process (more than one year), then the parameters $\gamma$ and $\theta$ cannot be considered as constants and will vary periodically. The period of the parameter $\gamma$ is $[0,360^{\circ}]$ and the parameter $\theta$ cycles from a minimum value of $66.5^{\circ}$ (autumnal equinox) to a maximum value of $113.5^{\circ}$ (vernal equinox), and their periods all correspond to one year. This suggests that the rate of change of the orbital elements in Eq.(\ref{eesc}) will have some periodically transformed oscillatory terms that can be neglected after averaging. With these considerations, the rates of change of the orbital elements are reduced to
\bqn
\dot p_3 &=& 0, \nb\\
\dot e_3  &=& -\f{3}{2}  \f{f'}{p} \sqrt{\f{GM_{\oplus}}{a^3}} v^2_{\oplus}  \sin \iota  \cos \iota \sin \omega, \nb\\
\dot \iota_3 &=& 0 , \nb\\
\sin \iota \dot \Omega_3  &=&  0 , \nb\\
\dot \omega_3 &=&   \f{3}{2}\f{f'}{ep} \sqrt{\f{GM_{\oplus}}{a^3}} v^2_{\oplus}   {  \sin \iota  \cos \iota \cos \omega } . 
\eqn
From the above expressions, we can clearly see that only the two orbital elements, eccentricity and pericentre, are truly non-zero. This indicates that the parity-violating effect does have an effect on the motion process of the satellite, which will gradually accumulate as the evolution time increases. { Meanwhile, we observe that the results for the semi-latus rectum,  inclination and ascending node of this theory agree with those in GR. It is shown that the parity-violating terms did not change the semi-latus rectum, while the eccentricity and ascending node of the orbital elements are affected only in a periodic manner, remaining unchanged after averaged over time. The only secular effects appeared in the eccentricity and pericentre of the orbital elements.}
Here we would not be able to use the observation of the ascending node in Eq.(\ref{lgu}) to constrain the parity-violating effect of the theory. 
It is worth noting that there is a clear difference between our computed ascending node advance and that in Ref.\cite{smith2008}, although both the CS gravity and the ghost-free parity-violating gravity have similar PPN parameters. The difference is that in Ref.\cite{smith2008} the parity-violating effect produces a correction for the ascending node while our result has no effect. This is due to the presence of some oscillatory terms in the solution of the vector Eq.(23) in Ref.\cite {smith2008}, which ensures the continuity of the solution at the sphere. These oscillatory terms are higher-order terms of the parameter $f'$ that are neglected when solving for the vector Eq.(4.15) in Ref.\cite{Qiao:2021fwi}.


We then turn to consider the measured pericentr advance of LAGEOS satellites around the Earth. Ref.\cite{Lucchesi:2010zzb} analyzed 13 years of tracking data from the LAGEOS satellites, providing a measurement of the pericentre advance of the LAGEOS satellite II as
\bqn
\dot \omega = \epsilon~ \dot \omega_{\rm GR} ,
\eqn
where $\epsilon= 1 + (0.28 \pm 2.14) \times 10^{-3}$ and the result of the LAGEOS satellite II for $\dot \omega_{\rm GR}$ has been given in Table. \ref{tableX}. This allows us to apply this observation to constrain the correction of the parity violation for the pericentr advance.
It is easy to derive a corresponding constraint as
\bqn
|\dot \omega_3| \lesssim |(\epsilon -1)\dot \omega_{\rm GR}|,
\eqn 
that can be translated into a constraint on the parameter $f'$ as
\bqn
f' \lesssim 10^4 ~{\rm m}.
\eqn
{The corresponding energy scale of the parity-violating in this gravity is $M_{\rm PV}:= 1/f' \gtrsim 10^{-20} {\rm GeV} $,  which is consistent with the constraint obtained from GPB\cite{Qiao:2021fwi}.
In non-dynamical CS gravity, the result of the measurement using the ascending node gives the parity-violating energy scale of $M_{\rm CS} \gtrsim 10^{-22} {\rm GeV} $, at which case the measurement precision considered is $10\%$\cite{smith2008}. Considering that the current measurement precision is already below $1\%$, the energy scale constraint obtained in non-dynamical CS gravity will be improved, whose result is compatible with ours.
In addition, a tighter constraint on the parity-violating energy scale in non-dynamical CS gravity based on measurements of the periastron precession rate of binary pulsar system is $M_{\rm CS} \gtrsim 10^{-18} {\rm GeV} $, which is two orders of magnitude higher than the results given in the solar system\cite{double_pulsar2}. Thus the contribution of the parity-violating effects will probably be more significant in binary pulsar systems, which will be considered in our future work. }

\begin{table*}
\caption{Orbital elements of the LAGEOS, LAGEOS II, and LARES satellites.}
\lb{table}
\begin{ruledtabular}
\begin{tabular} {cccccc}
Orbital elements & Unit & Symbol  &  LAGEOS  & LAGEOS II  & LARES\\
 \hline
semi-major axis & [km] & $a$ & 12270.00 & 12162.07 & 7820.31 \\
eccentricity & \;  &$e$ & 0.00451 & 0.01375 & 0.00074  \\
inclination & [deg]& $\iota$& 109.83  & 52.68  & 69.49\\
argument of perigee & [deg] & $\omega$ & 347.67 & 77.43 & 19.01 \\
right ascension of ascending node & [deg] & $\Omega$ & 249.38 & 26.14 & 294.48 \\
\end{tabular}
\end{ruledtabular}
\end{table*}

\begin{table}
\caption{Theoretical predictions of the secular precession rate on the right-ascension of the ascending node and pericenter of LAGEOS, LAGEOS II, and LARES satellites in general relativity. The unite of the rate is milliseconds of arc per year (mas/yr). }
\lb{tableX}
\begin{ruledtabular}
\begin{tabular} {cccccc}
Orbital elements &   LAGEOS  & LAGEOS II  & LARES\\
 \hline
$\dot \Omega_{\rm LT}$  & 30.68 & 31.50 & 118.50 \\
$\dot \omega_{\rm GR}$  & 3278.77+31.23  & 3352.58-57.31 & 10110.112-334.68 \\
\end{tabular}
\end{ruledtabular}
\end{table}

\section{Conclusions and Discussions}
\renewcommand{\theequation}{5.\arabic{equation}} \setcounter{equation}{0}

In this paper we extend the previous work by investigating the evolution of the motion of a binary system of Earth and its satellites in ghost-free parity-violating gravity. In this binary system, we consider both the influence of the Earth's spin and the translational motion on the evolution of the satellite orbital motion process. We applied the PPN metric of the ghost-free parity-violating gravity to calculated each orbital element of the satellite motion. 
Comparing the results with those of GR, we find that the parity violation terms correct for both eccentricity and pericenter in the all orbital elements, while the rest of the orbital elements are consistent with GR and conform to current experimental observations. These two corrections are secular effects that gradually accumulate over time. This suggests that these effects will eventually be tested experimentally, if observations are performed experimentally for the enough period of time. By using the observation of the pericenter advance of LAGEOS Satellite II, we give specific constraints on this theoretical parameter:
$ f' \lesssim 10^4 ~{\rm m}$.
We also give constraints on the theoretical parameters $f' \lesssim 10^4 ~{\rm m}$ in the Gravity Probe B experiment, with comparisons obtained in the LAGEOS Satellite II experiment found to be orders of magnitude the same.

It is expected that the LARRES$-2$ satellite has been launched on June 13, 2022, which joins the experiment and may enhance the detection capability of the experiment. As the experimental observations continue and cumulative data analysis improves, we expect observations of other orbital elements to be given, especially eccentricity. Once multiple orbital elements are available, we can use them to impose joint constraints on the parity-violating effects, which in turn give tighter constraints on the parameters.
In addition, we are more interested in testing the gravity with parity violation in binary pulsar systems in the future, which is another important testbed in the field of strong fields. Since the parity violation of gravity induces some effects on binary star systems that may be more significant, including modifications to the rate of periastron precession, the modification on the Schiff procession and/or the frame-dragging procession, as well as the modification on Lense-Thirring procession. Increasingly precise experimental observations of binary pulsar systems make this even more interesting.

\section*{Acknowledgements}

This work is supported by the National Natural Science Foundation of China (Grant No. 12325301 and 12273035), the National Key R\&D Program of China (Grant No. 2020YFC2201400 and 2021YFC2203100), and the Fundamental Research Funds for the Central Universities under Grant No. WK2030000036 and WK3440000004. Tao Zhu is partly supported by the National Key Research and Development Program of China Grant No.2020YFC2201503, the Zhejiang Provincial Natural Science Foundation of China under Grant No. LR21A050001 and LY20A050002, and the National Natural Science Foundation of China under Grant No. 12275238 and No. 11675143.

\appendix

\section*{Appendix A: PPN potentials}
\renewcommand{\theequation}{A.\arabic{equation}} \setcounter{equation}{0}
\lb{A11}
In this Appendix, we present the explicit expressions for the PPN potentials used to parameterize the metric in Eqs. (\ref{ppe1}). These potentials are given as follows \cite{will2018th}:
\bqn\lb{ppe2}
U & \equiv & \int \frac{\rho\left(\mathbf{x}^{\prime}, t\right)}{\left|\mathbf{x}-\mathbf{x}^{\prime}\right|} d^{3} x^{\prime}, \lb{SPU}\\
\Phi_{1} & \equiv & \int \frac{\rho^{\prime} v^{\prime 2}}{\left|\mathbf{x}-\mathbf{x}^{\prime}\right|} d^{3} x^{\prime} ,\\
\Phi_{2} & \equiv & \int \frac{\rho^{\prime} U^{\prime}}{\left|\mathbf{x}-\mathbf{x}^{\prime}\right|} d^{3} x^{\prime}, \\
\Phi_{3} & \equiv & \int \frac{\rho^{\prime} \Pi^{\prime}}{\left|\mathbf{x}-\mathbf{x}^{\prime}\right|} d^{3} x^{\prime}, \\
\Phi_{4} & \equiv & \int \frac{p^{\prime}}{\left|\mathbf{x}-\mathbf{x}^{\prime}\right|} d^{3} x^{\prime} \\
V_{j} & \equiv& \int \frac{\rho\left(\mathbf{x}^{\prime}, t\right) v_{j}^{\prime}}{\left|\mathbf{x}-\mathbf{x}^{\prime}\right|} d^{3} x^{\prime},\\
W_{j} & \equiv &\int \frac{\rho\left(\mathbf{x}^{\prime}, t\right) \mathbf{v}^{\prime} \cdot\left(\mathbf{x}-\mathbf{x}^{\prime}\right)\left(x-x^{\prime}\right)_{j}}{\left|\mathbf{x}-\mathbf{x}^{\prime}\right|^{3}} d^{3} x^{\prime}.\lb{SPchi}
\eqn

\section*{Appendix B: Osculating equations of Kepler problem with perturbations}
\renewcommand{\theequation}{B.\arabic{equation}} \setcounter{equation}{0}

The effects of the Lense-Thirring terms and the parity-violating effects  in the PPN  metric on the binary motion can be treated as perturbations to the Kepler problem in Newtonian mechanics. Under the perturbations, the equations of the osculating orbital elements are given by
\bqn
\frac{da}{dt} &=& 2 \sqrt{\frac{a^3}{G m}} (1-e^2)^{-1/2}  \times \Big[e \sin f \;{\cal R} + (1+e \cos f) {\cal S}\Big], \\
\frac{de}{dt} &=& \sqrt{\frac{a}{Gm (1-e^2)}} \times \left[ \sin f {\cal R} + \frac{2 \cos f + e (1+\cos^2f)}{1+e \cos f} {\cal S}\right], \\
\frac{d\iota}{dt} &=& \sqrt{\frac{a}{G m (1-e^2)}} \frac{\cos (\omega+f)}{1+e \cos f} {\cal W}, \\
\sin \iota \frac{d\Omega}{dt} &=& \sqrt{\frac{a}{G m(1-e^2)}} \frac{\sin(\omega+f)}{1+e \cos f} {\cal W}, \\
\frac{d \omega}{dt} &=& \frac{1}{e} \sqrt{\frac{a}{G m (1-e^2)}} \left[- \cos f {\cal R} + \frac{2 + e \cos f}{1+e \cos f} \sin f {\cal S} - e \cot \iota \frac{\sin (\omega+f)}{1+e \cos f} {\cal W}\right], \\
\frac{df}{dt} &=& \sqrt{\frac{G m (1-e^2)^3}{a^3}}(1+e \cos f)^2  + \frac{1}{e} \sqrt{\frac{a}{G m (1-e^2)}} \times  \left[\cos f {\cal R} - \frac{2 + e \cos f}{1+e \cos f} \sin f {\cal S}\right].
\eqn

In order to investigate the secular change of the orbital elements, we transform the derivatives with respect to $t$ in the above expressions to the ones with respect to the true anomaly $f$,
\bqn
\frac{da}{df} &\simeq & 2 \frac{p^{3}}{G m} \frac{1}{(1+e \cos f)^{3}} \mathcal{S}, \lb{af}\\
\frac{de}{df} & \simeq & \frac{p^{2}}{G m}\left[\frac{\sin f}{(1+e \cos f)^{2}} \mathcal{R} +\frac{2 \cos f+e\left(1+\cos ^{2} f\right)}{(1+e \cos f)^{3}} \mathcal{S}\right], \lb{ef}\\
\frac{d \iota}{df} &\simeq & \frac{p^{2}}{G m} \frac{\cos (\omega+f)}{(1+e \cos f)^{3}} \mathcal{W}, \lb{if}\\
\sin\iota \frac{d \Omega}{df}  &\simeq & \frac{p^{2}}{G m} \frac{\sin (\omega+f)}{(1+e \cos f)^{3}} \mathcal{W},  \lb{Omegaf}\\
\frac{d \omega}{df} & \simeq &
\frac{1}{e} \frac{p^{2}}{G m}\left[-\frac{\cos f}{(1+e \cos f)^{2}} \mathcal{R} +\frac{2+e \cos f}{(1+e \cos f)^{3}} \sin f \mathcal{S} -e \cot \iota \frac{\sin (\omega+f)}{(1+e \cos f)^{3}} \mathcal{W}\right]. \lb{omegaf}
\eqn

\end{document}